\documentclass[9pt,twocolumn,twoside]{revtex4}

\usepackage{bbm}
\usepackage{hyperref}
\usepackage{amsmath}
\usepackage{amsfonts}
\usepackage{amssymb}
\usepackage{graphicx}
\usepackage{empheq}
\usepackage{color}

\hypersetup{colorlinks=true, urlcolor=blue, linkcolor=blue, citecolor=blue,plainpages=false}

\newcommand{\txtpow}[1]{{\mbox{\scriptsize{#1}}}}
\renewcommand{\tensor}[1]{{\mathbbm{#1}}}
\newcommand{\myeqref}[1]{(\ref{#1})}

\begin{document}

\title{Whispering gallery mode single nano-particle detection and sizing: the validity of the dipole approximation}

\author{Matthew R. Foreman$^1$}
\email[]{matthew.foreman@imperial.ac.uk}
\author{David Keng$^2$}
\author{Eshan Treasurer$^2$}
\author{Jehovani Lopez$^2$}
\author{Stephen Arnold$^2$}
\email[]{sarnold935@aol.com}

\affiliation{$^1$Blackett Laboratory, Department of Physics, Imperial College London, London SW7 2AZ, UK\\
	$^2$Microparticle Photophysics Lab (MP$^3$L), NYU Polytechnic School of Engineering, Brooklyn, New York 11201, USA}
\date{\today}

\begin{abstract}
Interactions between whispering gallery modes (WGMs) and small nanoparticles are commonly modelled by treating the particle as a point dipole  scatterer. This approach is assumed to be accurate as long as the nanoparticle radius, $a$, is small compared to the WGM wavelength $\lambda$. In this article, however, we show that the large field gradients associated with the evanescent decay of a WGM causes the dipole theory to significantly underestimate the interaction strength, and hence induced WGM resonance shift, even for particles as small as $a\sim \lambda/10$. To mitigate this issue we employ a renormalized Born approximation to more accurately determine nanoparticle induced resonance shifts and hence enable improved particle sizing. The domain of validity of this approximation is investigated and supporting experimental results are  presented.
\end{abstract}

\maketitle

Nanoparticles, such as viruses, only exist in small concentrations in biological fluids.  This has compelled researchers to develop measurement techniques that can detect viruses and other particles at the ultimate sensitivity, i.e., one at a time. One such technique, utilising whispering gallery mode (WGM) micro-cavity transducers has proven to be a particularly sensitive and versatile platform for particle sensing and for studying the  interaction of nanoparticles with surface anchored antibodies \cite{Arnold2003,Sun2011a,Vollmer2008a,Foreman2015a}. WGM microcavities however also enable nanoparticle size to be measured through observation of the frequency shift \cite{Vollmer2008,Keng2014} or mode splitting \cite{Zhu2009,Kim2012e} that is induced when a nanoparticle binds to the resonator surface. With sufficient accuracy such size information can be used as a particle discriminant. 

Frequently, the WGM transduction mechanism is treated as an interaction between the WGM's evanescent near field and a point dipole induced in the nanoparticle, a model which is considered accurate so long as the particle radius, $a$, is small compared with the wavelength $\lambda$. In this article we show that even for small particles, dipole theory can underestimate the interaction strength resulting in potential sizing errors. This discrepancy arises from an inadequate description of the  field within the nanoparticle, which varies on the scale of the characteristic WGM decay length  as opposed to the wavelength. Using a renormalized Born approximation for the internal field we present analytic formulae which enable accurate particle sizing from WGM resonance shifts and hence overcome these limitations. Experimental results are presented to support our theory.

To get a feeling for the origin of the mode shift we start by describing the mechanism heuristically. As a nanoparticle enters the evanescent field of an unperturbed WGM generated by $\mathcal{N}$ trapped photons of frequency $f$, the field does reactive work $\Delta W$ to polarize the particle. The photons pay for this interaction by reducing their energy, $\mathcal{E}=\mathcal{N}hf$, generating a corresponding frequency shift $\Delta f$ in accordance with the polarization energy, $\mathcal{N}h\Delta f = -\Delta W$ \cite{Arnold2003}. The resulting fractional change in frequency is found by dividing the change in the energy by the mode energy, ${\Delta f}/{f} = -{\Delta W}/{\mathcal{E} }$. Importantly, this simple energy balance argument, known as the reactive sensing principle (RSP) \cite{Arnold2009}, is  consistent with first order perturbation theory \cite{Teraoka2006a}.  Since the polarization energy is related to the size of the perturbing particle, the measured frequency shift can be used for sizing \cite{Keng2014,Arnold2017}. 

\begin{figure}[!t]
	\begin{center}
		\includegraphics[width=0.95\columnwidth]{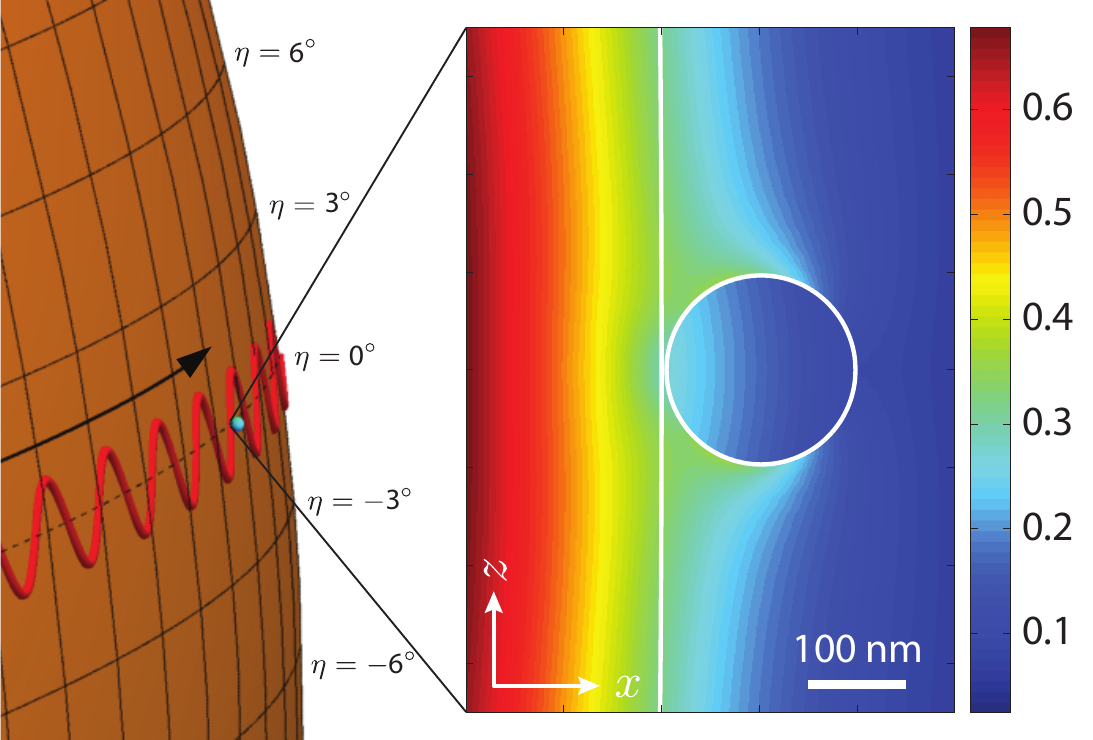}
		\caption{COMSOL calculation showing the mode distribution within a 96.7~nm radius polystyrene particle located at the equator of a spheroidal resonator and bathed in the field of a fundamental TE WGM at $\lambda = 1063$~nm. The position of the vertical white line represents the surface of the micro-cavity  \label{fig:1}}
	\end{center}
\end{figure}

Both the mode  and polarization energy can be written in terms of the WGM field distributions. 
The former, which is comprised of equal electric and magnetic energy contributions, can be expressed  solely in terms of the unperturbed electric field $\mathbf{E}(\mathbf{r})$ and is given by $\mathcal{E}  = (1/2) \int_V \epsilon(\mathbf{r}) |\mathbf{E}(\mathbf{r})|^2 dV$, where the integral is over the mode volume $V$ and $\epsilon(\mathbf{r})$ is the electric permittivity. On the other hand, the polarization energy depends on both the unperturbed and perturbed field ($\mathbf{E}'(\mathbf{r})$) according to $\Delta W  = (1/4) \mbox{Re}[\int_{V_p}\Delta \epsilon(\mathbf{r}) \mathbf{E}^*(\mathbf{r}) \cdot \mathbf{E}'(\mathbf{r}) dV ]$, 
where  $V_p$ denotes the volume of the particle and $\Delta\epsilon = \epsilon_p - \epsilon_m$ is the difference of the electric permittivity of the particle and  external medium. The fractional frequency shift can thus be written as \cite{Teraoka2006a}
\begin{align}
\frac{\Delta f}{f}= -\frac{\mbox{Re}\left[\int_{V_p}\Delta \epsilon(\mathbf{r}) \mathbf{E}^*(\mathbf{r}) \cdot \mathbf{E}'(\mathbf{r}) dV \right]}{2 \int_V \epsilon(\mathbf{r}) |\mathbf{E}(\mathbf{r})|^2 dV}. \label{eq:RSP}
\end{align}
Although the evanescent nature of the unperturbed field $\mathbf{E}$ is well known, it has become common to assume the particle is small enough that it can be treated as a point dipole positioned at the center of the particle $\mathbf{r}_p$. This approach is equivalent to assuming the sphere is illuminated by a uniform field thus producing a uniform field with magnitude $|\mathbf{E}'(\mathbf{r})| = 3\epsilon_m |\mathbf{E}(\mathbf{r}_p)|/(\epsilon_p + 2 \epsilon_m)$ within the particle. The familiar result \cite{Arnold2003}
\begin{align}
\frac{\Delta f_{\txtpow{dp}}}{f} = - \frac{\mbox{Re}[\alpha] |\mathbf{E}(\mathbf{r}_p)|^2}{2 \int_V \epsilon(\mathbf{r}) |\mathbf{E}(\mathbf{r})|^2 dV} \label{eq:RSP_dp}
\end{align}
subsequently follows, where $\alpha$, the dipole excess polarizability, is given by $\alpha = 4\pi \epsilon_m a^3 (\epsilon_p - \epsilon_m) / (\epsilon_p + 2\epsilon_m)$ for a spherical particle of radius $a$. A dipole model is quite appropriate for describing Rayleigh scattering of a plane wave by a particle whose radius is considerably smaller than the radiation wavelength $\lambda$. However, for near field problems, such as the interaction of the evanescent field of a WGM with a virus ($a \sim 100$~nm), the unperturbed evanescent intensity drops radially from the rim of the resonator to the center of the particle by $\exp[-a/L]$, where  the characteristic  intensity decay length, $L$, is considerably smaller than the wavelength $\lambda$. Such strong field gradients imply that use of a point dipole model can lead to substantial errors since higher order multipole contributions to the perturbed field are omitted \cite{Chaumet1998}. To illustrate the inaccuracy of assuming a uniform field we have performed finite element simulations using COMSOL, as detailed in \cite{Baaske2014}, for a micro-sphere resonator (albeit we note that our discussion also applies to other resonator geometries). Specifically, Figure~\ref{fig:1} shows the perturbed intensity distribution for a fundamental transverse electric (TE) WGM of order $l = 340$ excited at $\lambda \approx 1063$~nm in a spherical silica micro-cavity ($n_r = \sqrt{\epsilon_r} = 1.449$) with radius $R = 40.5$~$\mu$m and perturbed by an aqueous borne ($n_m = \sqrt{\epsilon_m} = 1.326$) polystyrene particle ($n_p = \sqrt{\epsilon_p} = 1.5719$, $a = 96.7$~nm) located at the equator.  Although $a/\lambda < 0.1$ we observe that the intensity within the particle falls off by a factor of $2.87$ over the  extent of the particle, closely matching the decay of $2.63$ of the unperturbed mode ($L=195$~nm) over the same distance. Use of the dipole theory is thus clearly inappropriate even for such a modestly sized nanoparticle.

To move beyond the limitations of the dipole model, we must evaluate \eqref{eq:RSP}, allowing for variation of the perturbed and unperturbed field distributions within the volume of the nanoparticle. Naturally, determination of the fields can be performed numerically using, for example, finite element \cite{Baaske2014}, mode matching \cite{Du2013} or boundary element methods \cite{Wiersig2003}, however, this approach can become computationally burdensome in particle sizing applications. Analytic formulae, as we develop below, are thus preferable in such cases. Although we make a number of approximations in our derivations, these are crucially less restrictive than those of the point dipole model.

Regardless of resonator geometry, the unperturbed WGM mode exhibits a rapid fall off of the mode intensity in the exterior volume of the resonator, which can be well approximated by an exponential decay \cite{Foreman2016, Demchenko2013}. Restricting to a spherical geometry for simplicity, we can thus write $\mathbf{E}(\mathbf{r}) = \mathbf{E}_0(\mathbf{R}) \exp[-\kappa  (r-R)]$, where $r=|\mathbf{r}|$ is the radial coordinate, $R$ is the resonator radius, $\kappa =  1/(2L)$ and  $\mathbf{E}_0(\mathbf{R})$ is the field at the cavity surface. Without loss of generality we assume that the particle is centered at $\mathbf{r}_p = (R+a)\hat{\mathbf{x}}$. Within any given cross-section of the nanoparticle, taken at a fixed axial ($x$) plane, the radial dependence of the unperturbed mode produces a smaller field amplitude at the nanoparticle surface relative to that at the center. Typically,  $a \ll R$ such that in the worst case the  amplitude ratio is $\approx \exp[\kappa a^2 / (R+a)] \approx 1$ even if $\kappa a \sim 1$. Variation of the polarization across the nanoparticle can also be similarly neglected. Consequently, the unperturbed mode within the nanoparticle  can be approximated as 
\begin{align}
\mathbf{E}(\mathbf{r}) = \mathbf{E}_0(R\hat{\mathbf{x}}) \exp[-\kappa (x-R)], \label{eq:E}
\end{align}
i.e., the WGM distribution is assumed constant for fixed $x$, however, the axial decay of the mode is still considered.

Determination of the perturbed field, however, requires a more in-depth analysis and in essence requires a solution of the electromagnetic scattering problem. As follows from the inhomogeneous vector wave equation, the perturbed mode within the nanoparticle is given by the self-consistent integral equation \cite{Novotny2006,Habashy1993}
\begin{align}
\mathbf{E}'(\mathbf{r}) = \mathbf{E}(\mathbf{r}) + \int\tensor{G}(\mathbf{r},\mathbf{r}')\Delta\epsilon(\mathbf{r}') \mathbf{E}'(\mathbf{r}') d\mathbf{r}' \label{eq:selfcons}
\end{align}
where $\tensor{G}(\mathbf{r},\mathbf{r}')$ is the dyadic Green's tensor of the system. After some algebraic manipulation Eq.~\myeqref{eq:selfcons} can be rewritten as
\begin{align}
\!\!\!\mathbf{E}'(\mathbf{r}) = \tensor{D}(\mathbf{r})\left[\mathbf{E}(\mathbf{r}) + \int\tensor{G}(\mathbf{r},\mathbf{r}')\Delta\epsilon(\mathbf{r}')( \mathbf{E}'(\mathbf{r}') - \mathbf{E}'(\mathbf{r})) d\mathbf{r}'\right]\label{eq:selfcons2}
\end{align}
where 
$\tensor{D}(\mathbf{r}) = [\tensor{I} - \int \tensor{G}(\mathbf{r},\mathbf{r}') \Delta\epsilon d\mathbf{r}']^{-1} \label{eq:depol}
$ 
is known as the depolarization tensor. Under the standard Born scattering approximation the field in the integral of Eqs.~\myeqref{eq:selfcons} and \myeqref{eq:selfcons2} is replaced by the incident (or unperturbed) mode distribution $\mathbf{E}(\mathbf{r}')$, however, the form of \eqref{eq:selfcons2} suggests the alternative approximation whereby the field in the integral is replaced by $\tensor{D}(\mathbf{r})\mathbf{E}(\mathbf{r})$. This is known in the literature as the renormalized Born approximation (RBA) or nonlinear localized approximation \cite{Habashy1993,Belkebir2005}. The error associated with making the RBA is given by the second term in \eqref{eq:selfcons2}. Noting that the Green's tensor possesses a strong singularity at $\mathbf{r} = \mathbf{r}'$ and assuming that it  falls off sufficiently rapidly away from this point, it follows that the dominant contribution to the integral in \eqref{eq:selfcons} arises from the field at $\mathbf{r}'=\mathbf{r}$. Accordingly, since the difference $\mathbf{E}'(\mathbf{r}') - \mathbf{E}'(\mathbf{r})$ is zero at this point it follows that the error term is small \cite{Habashy1993}. When the refractive index contrast of the scatterer  relative to the host medium, as parametrized by $\Delta\epsilon$, is small, the depolarization factor is approximately equal to the identity matrix such that the Born approximation is adequate. In contrast, when the refractive index difference is large the depolarization factor must be included.

\begin{figure}[!t]
	\begin{center}
		\includegraphics[width=\columnwidth]{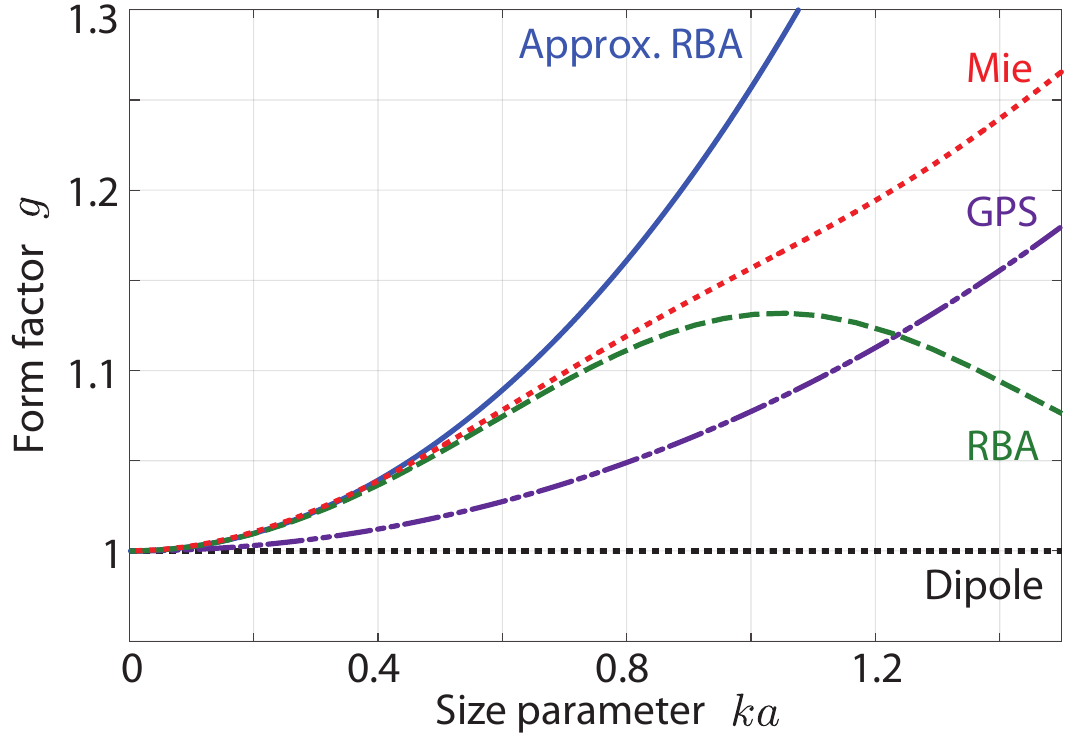}
		\caption{Variation of the WGM frequency shift relative to that of the dipole model, as calculated using the approximate analytic results of Eqs.~\myeqref{eq:g1} and \myeqref{eq:g2} (solid blue), numerical evaluation of Eqs.~\myeqref{eq:g1def} and \myeqref{eq:g2def} (dashed green), Eq.~(4) of \cite{Arnold2017} (dot-dashed purple), and Mie theory (dotted red). }\label{fig:gplots}
	\end{center}
\end{figure} 

Application of the above equations to spherical particles located at the origin in free space (i.e. neglecting any secondary scattering from the resonator surface) gives the internal perturbed field as
\cite{Habashy1993}
\begin{align}
\!\!\!\mathbf{E}'(\mathbf{r}) = \left[1-\frac{\Delta \epsilon}{\epsilon_m} h(\mathbf{r})\right]^{-1} \left[\mathbf{E}(\mathbf{r}) + \frac{\Delta\epsilon}{\epsilon_m} p(\mathbf{r}) \frac{
	\hat{\mathbf{r}} \cdot 
	\mathbf{E}(\mathbf{r})  \hat{\mathbf{r}} }{1- (\Delta\epsilon/\epsilon_m)s(\mathbf{r})  } \right] \label{eq:Eprime}
\end{align}
where we use caret notation to denote unit vectors, $k_m = n_m k$, $k = 2\pi/\lambda$, $s(\mathbf{r}) = p(\mathbf{r}) + h(\mathbf{r})$ 
\begin{align}
h(\mathbf{r}) &= -1 + \frac{\psi(k_m a)}{k_m r}\left[\sin(k_m r) + \frac{\cos(k_m r)}{k_m r} - \frac{\sin(k_m r)}{(k_m r)^2} \right] \\
p(\mathbf{r}) &= -\frac{\psi(k_m a)}{k_m r} \left[\sin(k_m r) + 3\frac{\cos(k_m r)}{k_m r} - 3\frac{\sin(k_m r)}{(k_m r)^2}\right]
\end{align}
and $\psi(z) = [1-iz] \exp[iz]$. Combining Eqs.~\myeqref{eq:RSP}, \myeqref{eq:E} and \myeqref{eq:Eprime} and accounting for the translation of the particle by defining $\mathbf{r}' = (x',y',z') = \mathbf{r} - \mathbf{r}_p$, allows us to determine the ratio, or form factor, $g= \Delta f/\Delta f_{\txtpow{dp}}$, which quantifies the resonance shift induced by a particle relative to the dipole model. Restricting  to non-absorbing media (i.e. real $\epsilon$) we find $g = g_1+g_2$ where
\begin{align}
g_1 &=\frac{1}{4 \pi a^3 }\int_{V_p} e^{-2\kappa x'}\left[\frac{\epsilon_p + 2\epsilon_m}{\epsilon_m -\Delta\epsilon h(\mathbf{r}')}\right]   d\mathbf{r}' \label{eq:g1def}\\ 
g_2 &=\frac{\Delta\epsilon}{4 \pi a^3 } \int_{V_p} e^{-2\kappa x'}\left[\frac{\epsilon_p + 2\epsilon_m}{\epsilon_m -\Delta\epsilon h(\mathbf{r}')}\right] \frac{  
	|\hat{\mathbf{r}'}\cdot\hat{\mathbf{E}}_0|^2 \,p(\mathbf{r}') }{\epsilon_m- \Delta\epsilon \,s(\mathbf{r}')  } d\mathbf{r}'. \label{eq:g2def}
\end{align}
In the small particle limit $a\rightarrow 0$, it can easily be shown that $h(\mathbf{r}')\rightarrow -1/3$ and $p(\mathbf{r}')\rightarrow 0$, such that $g\rightarrow 1$ as would be expected.
Practically, these integrals must be evaluated numerically, however, by expanding the kernels with respect to $a$ and $r'$ up to second order we can derive an approximate closed form expression for $g$. For $g_1$ we have 
\begin{align}
g_1 \approx \frac{3}{4 \pi a^3 }\int_{V_p} e^{-2\kappa x'}\left[1 + \frac{ k^2 a^2\alpha}{4\pi a^3} - \frac{k^2 r'^2 \alpha}{10 \pi a^3}\right] d\mathbf{r}' .
\end{align}
Letting $\zeta = 2 \kappa a$ this integral can be simply evaluated yielding
\begin{align}
g_1 &\approx \frac{3}{\zeta^3}\left[\zeta \cosh\zeta - \sinh \zeta\right] \nonumber \\
&\quad+ \frac{3\alpha}{4\pi a^3} \frac{k^2 a^2}{5\zeta^5}\left[3\zeta(\zeta^2 - 4)\cosh\zeta + (\zeta^2 + 12)\sinh\zeta\right].\label{eq:g1}
\end{align}
We note that the first term of \eqref{eq:g1} (henceforth denoted $g_{\txtpow{GPS}}$) corresponds to the form factor reported in \cite{Arnold2017}. To determine $g_2$ we first express the polarization dependent term, $|\hat{\mathbf{r}}'\cdot\hat{\mathbf{E}}_0|^2$, using polar and azimuthal angular coordinates taken relative to the center of the nanoparticle, before expanding the kernel of $g_2$ up to and including quadratic terms, whereby we find
\begin{align}
g_2 &\approx \frac{3 \alpha}{4\pi a^3}\frac{k^2 a^2}{5 \zeta^5} \left[(\zeta^2 + 3)\sinh\zeta - 3 \zeta \cosh\zeta \right.\nonumber\\
&\quad\left.+ |e_x|^2\left\{\zeta(\zeta^2 + 15)\cosh\zeta - 3(5 + 2\zeta^2)\sinh\zeta\right\}\right]. \label{eq:g2}
\end{align}
As discussed in Ref. \cite{Foreman2016}, $|e_x|^2$ is zero for TE modes and $\approx \epsilon_r / (2\epsilon_r-\epsilon_m) $ for transverse magnetic (TM) WGMs. Within the RBA we find that the polarization dependent $|e_x|^2$ term in \eqref{eq:g2} plays a negligible role, such that in simulations we restrict attention to TE modes. Figure~\ref{fig:gplots} illustrates the variation of $g^{\txtpow{app}}_{\txtpow{RBA}}$, as follows from Eqs.~\myeqref{eq:g1} and \myeqref{eq:g2},  with nanoparticle size (solid blue curve) as compared to the dipole model for which $g_{\txtpow{dp}}=1$ by definition (dotted black). Simulation parameters are the same as those used in the finite element calculations discussed above. Results from numerical evaluation of $g_{\txtpow{RBA}}$, as given by Eqs.~\myeqref{eq:g1def} and \myeqref{eq:g2def}, are also shown (dashed green curve). Additionally we have used Mie theory to determine the mode distribution within the nanoparticle when illuminated by an evanescent wave in a total internal reflection configuration, including surface dressing effects \cite{Frezza2015}. Calculated mode distributions were then subsequently used to evaluate the RSP integral (\eqref{eq:RSP}), and hence $g_{\txtpow{Mie}}$, numerically which is shown in Figure~\ref{fig:gplots} (dotted red curve). Good agreement between $g_{\txtpow{Mie}}$ and the approximate RBA form factor, $g^{\txtpow{app}}_{\txtpow{RBA}}$, up to size parameters of $ka \approx 0.6$ is seen, at which point the relative error is approximately 1\%, whereas it is 7.2\% for the dipole model. The full RBA form factor $g_{\txtpow{RBA}}$ suffers from only a 0.3\% relative error for particles of this size. At $ka = 1$ these errors increase to 8.7\%, 13.5\%, and 2.2\% respectively. Finally the purple dot-dashed line shows variation of $g_{\txtpow{GPS}}$  (corresponding to the first term of \eqref{eq:g1}). Whilst it is seen that this performs much better than the dipole approximation at larger particle sizes, it under performs with respect to the RBA results. Unphysical oscillations in $g_{\txtpow{RBA}}$ arise for $ka \gtrsim 1$ due to the approximations made and hence we do not apply the RBA to particle sizes larger than this limit in what follows.

\begin{table}[!t]
	\begin{tabular}{l|c|c|c}		
		$\langle a \rangle_{\txtpow{DCP}}$ (nm)& $94.5 \pm 2.0$ & $177.0 \pm 1.4$  & $220.5 \pm 1.8$  \\ \hline\hline
		$N$ & 26 & 24 & 15 \\
		$k\langle a\rangle_{\txtpow{DCP}}$  & 0.56 & 1.05 & 1.30 \\ \hline
		$\langle a \rangle_{\txtpow{dipole}}$   (nm)& $\color{red} {98.2 \pm 1.1}$ & $\color{red}{187.9 \pm 1.8}$& $\color{red}{248.5 \pm 2.9}$ \\
		$\langle a \rangle_{\txtpow{GPS}}$  (nm)& $\color{blue}97.2 \pm 1.1$ & $\color{red}180.2 \pm 1.6$ & $\color{red} 229.8 \pm 2.3$  \\
		$\langle a \rangle_{\txtpow{RBA}}$  (nm)& $\color{blue}95.6 \pm 1.1$ & $\color{blue}176.9 \pm 1.6$  & $-$  \\
		$\langle a \rangle_{\txtpow{Mie}}$  (nm)& $\color{blue}95.5 \pm 1.1$ & $\color{blue}174.6 \pm 1.5$ & $\color{blue}{222.4 \pm 2.1}$ \\ 
		 $\langle a \rangle_{\txtpow{RBA}}^{\txtpow{app}}$   (nm)& $\color{blue}95.2 \pm 1.1$ & $\color{red}{168.4 \pm 1.3}$ & $-$ 
	\end{tabular}\caption{		Comparison of the mean radii of three different polystyrene hydrosol ensembles as found through DCP ($\langle a \rangle_{\txtpow{DCP}}$) and WGM  sizing measurements ($\langle a \rangle_{g}$). Analysis of WGM data was performed using 5 different theoretical form factors $g$. $N$  measurements were performed for each ensemble with a standard deviation of $\sigma$. Mean radii are reported along with the expected standard deviation of the mean $ \sigma/\sqrt{N}$. Blue (red) numbers show results in (dis)agreement with  $\langle a \rangle_{\txtpow{DCP}}$.
\label{tab:results}}
\end{table}

To further test whether point dipole theory provides an adequate description for accurate particle sizing we have performed sizing measurements on three sets of particles lying near and beyond the edge of the Rayleigh regime. Additionally, we compare the accuracy of the other theoretical approaches described above (see Figure~\ref{fig:gplots}), which account for the finite size of the nanoparticle through the differing form factors $g$. Reference sizes were determined using Disc Centrifuge Photosedimentometry (DCP). Seven ensemble measurements were taken for each particle size yielding mean radii of $\langle a \rangle_{\text{DCP}} = 94.5
\pm 2.0$, $177.0\pm 1.4$ and $220.5\pm 1.8$~nm or, equivalently, optical sizes ($ka$) of 0.57, 1.05 and 1.30 at $\lambda = 1063$~nm. The core of the experimental setup for WGM particle sizing measurements comprised of a micro-spheroid resonator fabricated by melting the end of a tapered silica optical fiber using a CO$_2$ laser. Shape analysis of images of the resonators revealed that they were slightly prolate ($(R_p - R_e)/(R_p R_e^2)^{1/3} < 0.03$, where $R_p$ is the polar radius and $R_e$ is that of the equator). A slight eccentricity is required in our approach, full details of which can be found in Refs. \cite{Keng2014} and \cite{Arnold2017}, so as to lift the degeneracy of WGMs of different polar order $m$. The equatorial radius of each of the resonators varied from 40.5 to 43.5~$\mu$m with each radius measured to better than $\pm 1\%$. Resonators were immersed in a microfluidic cell containing a NaCl salt solution (between 20 and 30~mM at neutral pH) and over-coupled to a tapered optical fiber, in order to excite WGMs in the resonator propagating with the same sense. Nano-particles were subsequently injected into the cell. A tunable distributed feedback laser (DFB) coupled into the fiber was used to monitor the free space wavelengths of the $m=l$ and $m=l-1$ resonances. Steps in the resonance wavelength of these modes were recorded as particles bound to the resonator surface.  By taking the ratio of the measured shifts the latitude of the nanoparticle was then determined, which when combined with the RSP including any relevant form factor $g$, enabled the particle size to be determined from a single binding event. All of the form factors $g$ shown in Figure~\ref{fig:gplots} were used to analyse our experimental data. Since we consider only the frequency shift sizing modality we took special care to ensure that the induced WGM shift was considerably smaller than the resonance line-width. In total $65$ binding events were recorded (26 for the particles with $\langle a \rangle_{\text{DCP}} = 94.5$~nm, 24 for the $\langle a \rangle_{\text{DCP}} = 177.0$~nm particles and 15 for the $\langle a \rangle_{\text{DCP}} = 220.5$~nm particles). The results of our analysis are  listed in Table~\ref{tab:results}.

We note that for all particle sizes investigated, application of dipole theory to our micro-cavity experiments leads to nanoparticle sizes in excess of the DCP results. This disparity is due to the relatively small interaction strength associated with the dipole model. This discrepancy is most apparent for the 177.0~nm and 220.5~nm particles for which use of the dipolar $g$ factor yields mean radii of 187.9~nm and 248.5~nm. As one moves down the rows in Table~\ref{tab:results} the strength of the reactive interaction increases, and therefore the inferred nanoparticle size decreases. Approximate RBA theory clearly underestimates the radius of the 177~nm particles by nearly 9~nm, although performs adequately for the smaller 94.5~nm radius particles. The full RBA theory appears to provide good agreement for particles with DCP radii of 94.5~nm and 177.0~nm, although it is difficult to separate $\langle a \rangle_{\txtpow{GPS}}$, $\langle a \rangle_{\txtpow{RBA}}$, and $\langle a \rangle_{\text{Mie}}$ for the smaller particles due to statistical uncertainties in the data. For particles with size parameter greater than unity, the RBA theory breaks down and only full Mie theory calculations, including surface dressing effects, can yield correct particle sizes.

In summary, we have demonstrated both theoretically and experimentally that the commonly used dipole scattering approximation is inappropriate when considering the interaction between the evanescent field of a WGM and a small nanoparticle. Accurate particle sizing necessitates the decay of the WGM across the particle to be accounted for. We have presented more accurate expressions for the particle induced resonance shift based upon use of the RBA. These were found to enable accurate particle sizing for size parameters of $ka \lesssim 1$. Yet larger particles were found to require more rigorous electromagnetic modelling techniques to produce satisfactory sizing results.

\textbf{Funding.} MRF is funded by a Royal Society University Research Fellowship, and work at NYU by the US National Science Foundation grant EECS 1303499.

\end{document}